\documentclass[10pt,journal,compsoc]{IEEEtran}
\ifCLASSOPTIONcompsoc
  \usepackage[nocompress]{cite}
\else
  \usepackage{cite}
\fi
\usepackage[pdftex]{graphicx}

\usepackage{algorithm}
\usepackage{algpseudocode}
\usepackage{amsmath}

\usepackage{color}
\definecolor{ForestGreen}{RGB}{162,52,0}

\usepackage{xcolor}
\usepackage{amsmath}
\usepackage{CJK}
\usepackage{url}
\usepackage{array}

\usepackage{ragged2e}

\ifCLASSINFOpdf
\else
\fi

\begin{document}
%
\title{A Reliable Data-transmission Mechanism using Blockchain in Edge Computing Scenarios}

\author{Peiying Zhang, Xue Pang, Neeraj Kumar, \IEEEmembership{Senior Member,~IEEE}, Gagangeet Singh Aujla, \IEEEmembership{Senior~Member,~IEEE}, Haotong Cao, \IEEEmembership{Member,~IEEE}
\IEEEcompsocitemizethanks{\IEEEcompsocthanksitem This work is partially supported by the Major Scientific and Technological Projects of CNPC under Grant ZD2019-183-006, and partially supported by "the Fundamental Research Funds for the Central Universities" of China University of Petroleum (East China) under Grant 20CX05017A, 18CX02139. \textit{(Corresponding authors: Peiying Zhang and Neeraj Kumar.)}
}

\IEEEcompsocitemizethanks{\IEEEcompsocthanksitem P. Zhang and X. Pang are with the College of Computer Science and Technology, China University of Petroleum (East China), Qingdao 266580, China. (E-mail: zhangpeiying@upc.edu.cn, 1103746978@qq.com)
}

\IEEEcompsocitemizethanks{\IEEEcompsocthanksitem N. Kumar is with the Department of Computer Science and Engineering, Thapar Institute of Engineering and Technology. N. Kumar is also with Department of Computer Science and Information Engineering Asia University Taiwan and King Abdul Aziz University Jeddah Saudi Arabia. (Email: neeraj.kumar@thapar.edu)
}

\IEEEcompsocitemizethanks{\IEEEcompsocthanksitem G. S. Aujla is with the School of Computing, Newcastle University, United Kingdom. (E-mail: gagi\_aujla82@yahoo.com)
}

\IEEEcompsocitemizethanks{\IEEEcompsocthanksitem H. Cao is with the Jiangsu Key Laboratory of Wireless Communications, Nanjing University of Posts and Telecommunications, Nanjing 210003, P.R. China. (Email: caohaotong@163.com)
}

\thanks{ }}

%
%

\markboth{IEEE Internet of Things Journal}%
{Shell \MakeLowercase{{et al.}}: A Reliable Data-transmission Mechanism using Blockchain in Edge Computing Scenarios
}

\IEEEtitleabstractindextext{%
\begin{abstract}
\justifying\let\raggedright\justifying
With the advent of the Internet of things (IoT) era, more and more devices are connected to the IoT. Under the traditional cloud-thing centralized management mode, the transmission of massive data is facing many difficulties, and the reliability of data is difficult to be guaranteed. As emerging technologies, blockchain technology and edge computing (EC) technology have attracted the attention of academia in improving the reliability, privacy and invariability of IoT technology. In this paper, we combine the characteristics of the EC and blockchain to ensure the reliability of data transmission in the IoT. First of all, we propose a data transmission mechanism based on blockchain, which uses the distributed architecture of blockchain to ensure that the data is not tampered with; secondly, we introduce the three-tier structure in the architecture in turn; finally, we introduce the four working steps of the mechanism, which are similar to the working mechanism of blockchain. In the end, the simulation results show that the proposed scheme can ensure the reliability of data transmission in the Internet of things to a great extent.
\end{abstract}

\begin{IEEEkeywords}
Internet of Things, Data-transmission Mechanism, Blockchain, Edge Computing
\end{IEEEkeywords}}

\maketitle
\IEEEdisplaynontitleabstractindextext
\IEEEpeerreviewmaketitle

\section{Introduction}
The IoT is a large-scale network formed by the integration of various wired and wireless devices. Relevant research shows that by 2025, there will be more than 60 billion devices connected to the IoT, and the generation of massive data will make the IoT face great challenges\cite{DBLP:journals/iotj/PyoungB20}\cite{DBLP:journals/access/DingNRC20}. There are various types of sensors in the IoT. Each sensor, as an information source, can receive and send data to each other, which makes the IoT become a comprehensive system integrating information collection and processing.

\begin{figure*}[htbp] 
\centering
\includegraphics[width=2.0\columnwidth]{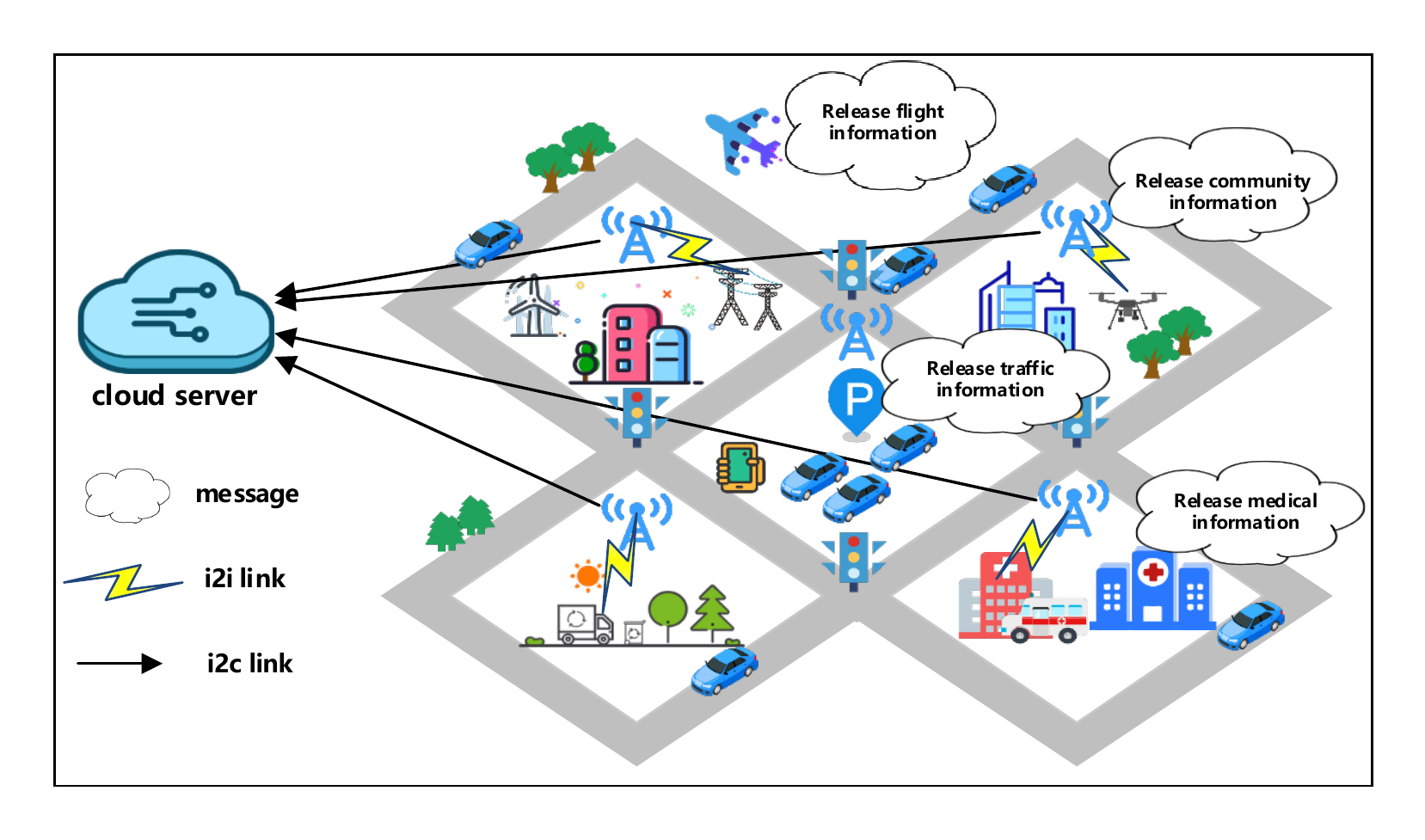}
\caption{Overview of transmission in the IoT.}
\label{iot}
\end{figure*}

With the rapid increase of devices connected to the IoT, data processing, dissemination, and collection become more and more common\cite{DBLP:journals/corr/abs-2005-01346}\cite{DBLP:journals/comsur/StoyanovaNPPM20}\cite{DBLP:journals/access/DianVR20}\cite{DBLP:journals/eswa/BamakanMB20}. Fig. \ref{iot} shows various scenarios of data transmission in the traditional IoT, including data transmission in the community, data transmission in the Internet of vehicles, data transmission in medical system, data transmission in flights, etc. The black arrow in the figure represents the transmission from infrastructure to cloud (i2c), and the yellow arrow represents the transmission from infrastructure to infrastructure (i2i).
It brings many problems \cite{DBLP:journals/corr/abs-2005-12685}, especially how to ensure the reliability of data. However, there are still many risks and challenges to solve the reliability of data transmission in the IoT:

(1) At present, the architecture of the IoT mainly adopts the mode of IoT cloud, which takes the cloud as the general agent of the center, that is, all the data are centralized to the cloud for processing and distribution. However, once this centralized processing mode is attacked illegally, the whole system will be paralyzed, which will cause catastrophic data loss and tampering.

(2) The centralized data processing method is difficult to expand, but now the data of the IoT is facing the explosive growth, and the cloud server will be difficult to expand in a short time. The influx of large quantities of data makes the cloud overload operation, unable to guarantee the integrity and timeliness of the data. For security aware and time-delay sensitive applications, the It would be a huge loss.

(3) In the process of data transmission, if the data is first transmitted to the cloud, and then sent to the IoT after processing, it will cause data leakage, and it is difficult to ensure user privacy. Therefore, it will make it difficult for some IoT applications to provide high-quality services.

In the future research, blockchain technology is the favorite of the academic community, and it will shine in various fields\cite{DBLP:journals/csur/LaoLHXGY20}\cite{DBLP:journals/access/ChoiL20}. In the IoT, blockchain technology can solve the problems caused by centralized data processing to a large extent, such as enhancing the reliability of data transmission, protecting the privacy of users so that data does not leak, so as to ensure the security of the Internet of things. In short, blockchain can be seen as a distributed bill, which was originally proposed by Nakamoto for bitcoin encryption. Because blockchain records data point-to-point and distributed, its data can hardly be tampered with.

EC technology has made a great innovation in the IoT technology. In the IoT, all devices are connected with cloud devices, and all data must be processed by cloud server. This not only increases the amount of data transmission, but also easily brings security problems such as data leakage. EC technology connects the IoT devices with the adjacent edge devices. Most of the information can be directly processed by edge devices, which greatly reduces the interaction between cloud servers and devices, and improves the security.

In this paper, we combine the characteristics of EC and blockchain to solve the reliability problem of data transmission in the IoT. Specific contributions can be summarized as follows:

(1) First of all, we propose a data transmission mechanism based on blockchain. It abandons the traditional centralized management mode, removes the intermediate agent, and adopts the block chain distributed framework with immutable data. The whole architecture is divided into three layers, which ensures that the data transmitted by users is not tampered with, improves the reliability, and protects the privacy of data users from being disclosed.

(2) Considering that when transmitting data between various sensor facilities in the IoT, there may be malicious users who destroy the order of the system, we define a mechanism similar to proof of credit, which can calculate the credit value of each participant, and the user with low credit can not participate in data transmission.

(3) We introduce the four working steps of the mechanism, which are registration process, data transmission process, bookkeeper selection process and reward process. Each process is proposed based on the working mode of blockchain, and the combination of edge computing can better ensure the reliability of data transmission.

The remainder of this paper is organized as follows. Section 2 reviews the existing methods for IoT and blockchain. Section 3 introduces the design of framework. In Section 4, we describe our proposed method in detail. The performance of our method and other methods is evaluated in Section 5. Section 6 concludes this paper.

\section{Related works}
In this chapter, we review some existing papers \cite{DBLP:journals/iot/Minoli20a}\cite{DBLP:journals/jsac/JiangWLWW20}\cite{DBLP:journals/iotj/BiswasSLMMW20}, which have made a lot of discussions in the field of blockchain and IoT \cite{DBLP:journals/tnse/JiangWSS20}\cite{DBLP:journals/fgcs/GianiniFMCBD20}\cite{DBLP:journals/tii/JiangWLQS20}. Next, we will introduce some representative works.

The authors of \cite{DBLP:journals/adhoc/MostefaouiFN19}\cite{DBLP:journals/jpdc/MerzougBMC19} puts forward several methods to solve the problems encountered in the IoT. Based on the blockchain technology, the authors of document \cite{DBLP:journals/iotj/Novo18} propose an architecture to manage the IoT devices, which connects the sensor devices in the IoT into the decentralized control system, avoiding the original centralized control. In order to solve the security problem of vehicle network physical system, the authors of \cite{DBLP:journals/fgcs/BaliK16} propose a novel security cluster for effective data distribution between different devices in vehicle network.

The authors of literature \cite{DBLP:phd/basesearch/Dorri20} and \cite{DBLP:journals/sp/WangLPLWZ20} propose three blockchain systems that are suitable for combination with the IoT. The first system is an extensible lightweight blockchain system, which makes great progress in reducing latency and overhead. The second system is a memory optimized and flexible blockchain system, which can greatly reduce memory occupation. The third system is used for transaction And reward flexible model. In literature \cite{DBLP:journals/jms/AminIBKK15}\cite{DBLP:journals/cee/ChallaDOKKKV18}, in order to solve the network insecurity of medical system, the authors designed a medical system, which can pass the standard authentication scheme, which is the security protection of patients' information.

The authors of literature \cite{8989000} combine blockchain with Virtual Network Function (VNF), and regard all processes related to VNF, such as instantiation, upgrading, retrieval, etc., as transactions in blockchain, and stores these transactions in a distributed ledger that will never change. Therefore, this paper proposes a decentralized framework, which is a P2P network composed of nodes of optical network, so as to realize the communication of optical network\cite{DBLP:journals/jsac/JiangCLR13}.

In document \cite{DBLP:journals/cm/Rawat19}, the authors introduce three new technologies: SDN, EC and blockchain technology, and propose a solution of wireless network virtualization by integrating these three technologies. This scheme can reduce the collision between these services and improve the inter operability of the industry. The authors of reference \cite{DBLP:journals/cn/JindalAK19} combine SDN with blockchain and propose an edge as a service framework based on blockchain.

At present, the cloud centric IoT brings about the delay problem, which hinders the interaction between sensors. The authors of literature \cite{DBLP:conf/IEEEcit/SamaniegoD16} propose a concept of micro-service, which analyzes the performance of virtual resources in different environments. The authors of literature \cite{DBLP:journals/fgcs/LuXLWZZ19} and \cite{DBLP:conf/icsa/WeberLTDGS19} have done research on the platform architecture based on blockchain. In literature \cite{DBLP:conf/icsa/WeberLTDGS19}, the authors propose an extensible platform architecture for the system based on multi tenant blockchain to ensure data integrity, while maintaining data privacy and performance isolation.

Autonomous driving is a very complex system, which closely integrates many technologies, and the smooth interaction with cloud platform is also an important issue. These problems bring many challenges to the development of automatic driving technology. Including data processing, load balancing, security threats, etc. The authors of literature \cite{DBLP:journals/pieee/LiuLTYWS19} summarize the existing methods of automatic driving, and further discuss them.

In recent years, many researches combine the industrial IoT with the blockchain, which is novel\cite{DBLP:journals/twc/JiangC0L13}. However, the blockchain has a high demand for resources, and the resources of the industrial IoT are very limited, which brings challenges to the research. In literature \cite{DBLP:journals/tii/LiuWLX19}, the authors propose a lightweight blockchain system, which can improve resource utilization and simplify complex formulas and concepts. It not only makes the block generation faster, but also reduces the memory consumption and greatly improves the system performance.

Point to point energy trading is widely used in the current industrial IoT, but the opaque trading market will bring many security risks\cite{DBLP:journals/jsac/ZhuJKGL17}. As a serious problem, security has received extensive attention in the academic community. In document \cite{DBLP:journals/tii/LiKYYDZ18}, the authors use blockchain technology to propose a safe energy trading system, which can reduce the delay of trading and enhance the trust of trading when trading point-to-point. Finally, the contribution of the system to security in industrial blockchain is proved.

The authors of document \cite{DBLP:journals/network/AujlaSBKHB20} propose a blockchain service framework for SDN, which combines the advantages of SDN and blockchain, and can solve the problem of secure transmission between various applications in smart city.

Because the IoT is distributed in all aspects of life, the authors of literature \cite{DBLP:journals/tii/ZomayaKRA20} introduce many applications of the IoT. In document \cite{DBLP:conf/ic2e/LiJAMZR20}, an IoT workflow composite system is proposed, which enables IoT users to pipeline their workflow in the proposed IoT workflow activity abstraction mode.

In document \cite{DBLP:journals/access/LoLCXLZN19}, the authors analyze various solutions for blockchain and IoT, and review the articles on these solutions. The authors analyze the problems in the development of IoT from two aspects of data and things, and study the solutions.

\section{The architecture of Internet of things based on blockchain}
\begin{figure}[htbp] 
\centering
\includegraphics[width=1.0\columnwidth]{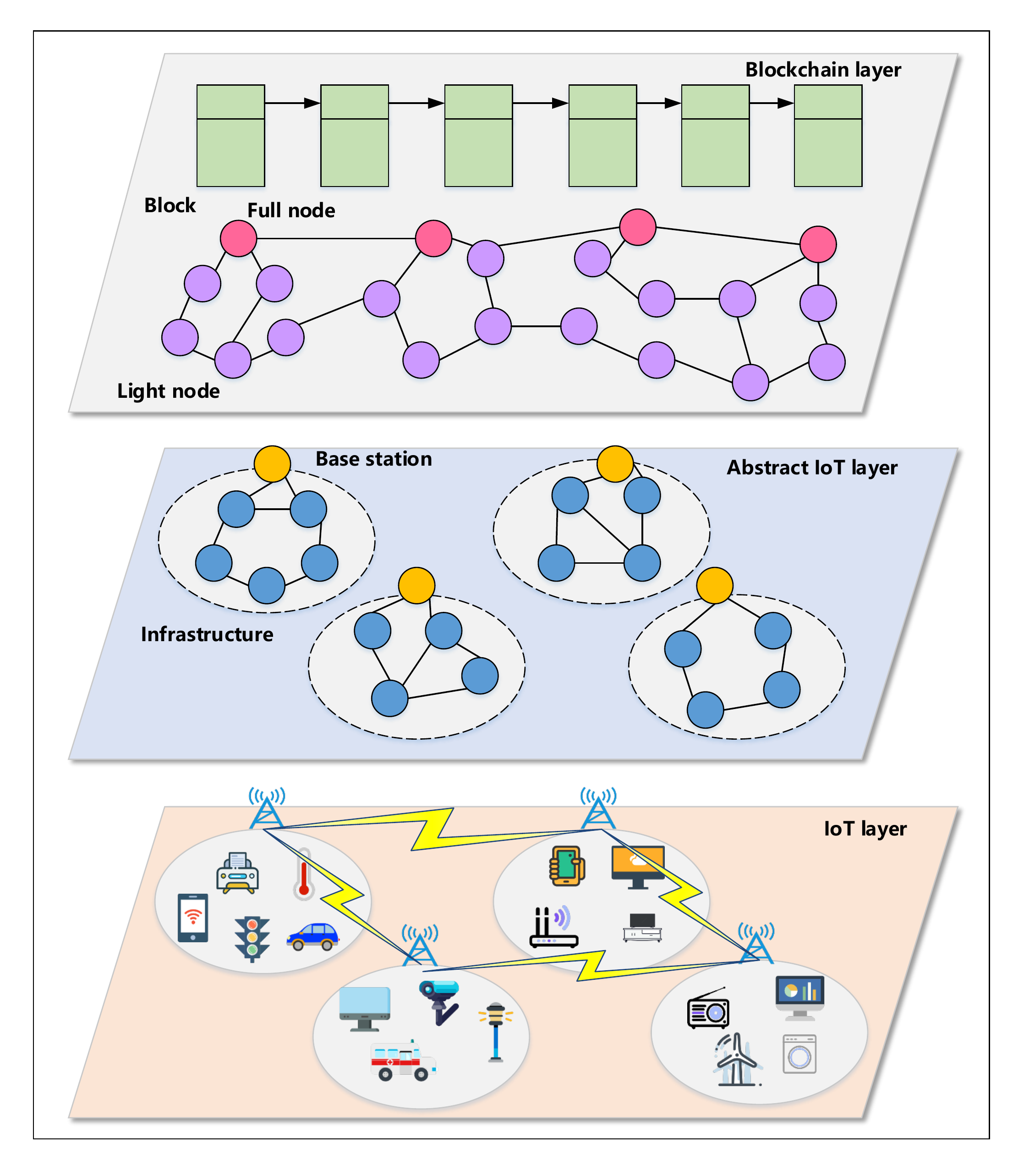}
\caption{Overview of the proposed architecture.}
\label{2}
\end{figure}

In this chapter, we introduce the IoT architecture based on blockchain proposed in this paper. The architecture of the IoT based on the blockchain is shown in Fig. \ref{2}. The architecture consists of three layers: the IoT layer, the abstract IoT layer and the blockchain layer. The following three levels are described in detail.

IoT layer: The IoT layer is a network layer composed of various physical infrastructures, which is located at the bottom of the architectures. As can be seen from the figure, the IoT includes a large number of infrastructures, including physical facilities with sensor equipment and edge nodes. According to the geographical location of each device, it can be connected with adjacent edge nodes and managed by edge nodes. The devices managed by each edge node can form a cluster, and the devices in each cluster have adjacent geographical location and similar demand environment. When the physical device needs to send information, it needs to request the edge node first. After the edge node verifies its security, it packages the information into the block and distributes rewards for it.

Abstract IoT layer: We abstract the IoT into a physical network model. IoT devices are represented by physical nodes, and edge nodes can also be represented as the IoT. However, two nodes are labeled with different colors, representing different types of devices. The cluster managed by each edge node can be represented by a physical domain.

Blockchain layer: Before joining the blockchain layer, all nodes need to be registered and authorized. In the blockchain environment, nodes can be divided into two types, full node and light node. Full nodes are played by edge nodes, which are responsible for managing all information and verifying every transaction. Light nodes are played by ordinary physical nodes, which do not participate in the construction and maintenance of the blockchain, but use the information in the blockchain for their own activities. When the information requested by the optical node passes the security verification of the whole node, the whole node will package the information into the block. The packaged blocks will be connected to the blockchain, and the connection will yield benefits. Finally, dividends will be paid according to the contribution. The top of the blockchain layer in Fig. \ref{2} shows a simple blockchain model. In the blockchain, each block is composed of two parts. The first part is the block header, which stores the hash value, version number, hash value of the previous block, Merkle tree root hash value and other information. The second part is the block body, recording the transaction list. Each block is linked by a hash pointer.

\section{Work process of the model}
\begin{figure*}[htbp] 
\centering
\includegraphics[width=2.0\columnwidth]{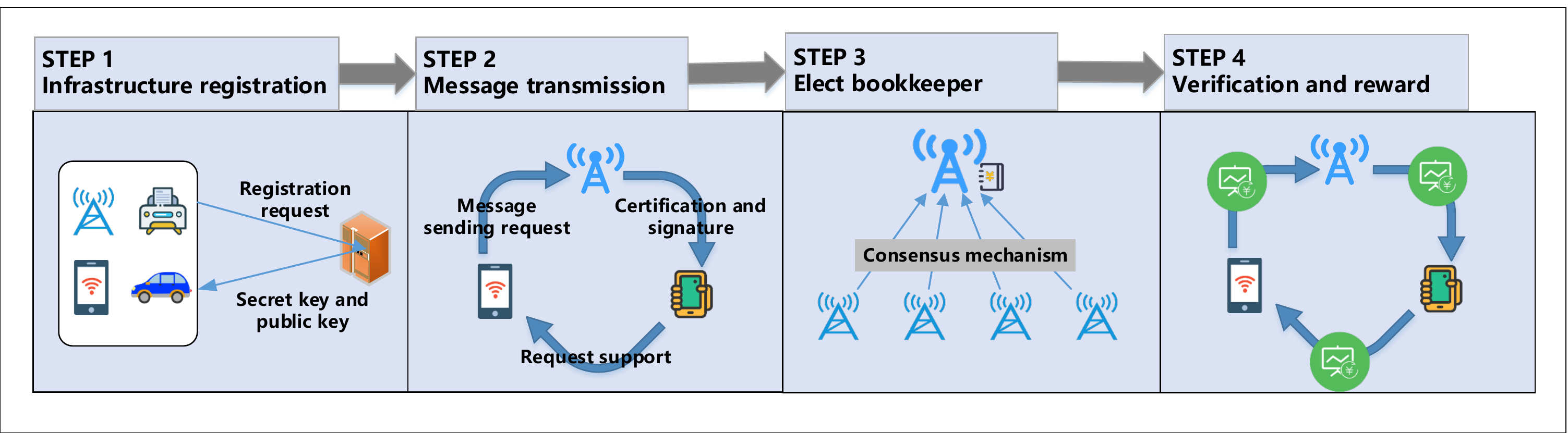}
\caption{Workflow chart of the framework.}
\label{3}
\end{figure*}

In this chapter, we introduce the operation process of the architecture. Generally speaking, it can be divided into four major steps. Firstly, the infrastructure in the IoT needs to be registered. Secondly, the infrastructure with sensors communicates through the edge nodes. Thirdly, the edge nodes use the formula protocol to select bookkeepers, package all information, connect to the blockchain. Fourthly, verify the reliability of the message and reward it. Fig. \ref{3} shows the whole process.

\subsection{Infrastructure registration}
When an infrastructure with sensors wants to send messages within the framework, it must be registered and authorized. Registration occurs when the infrastructure first enters the network. First, the infrastructure sends a registration request to the level agent, which includes the user identification information of the infrastructure, such as access code, equipment number, etc. After the initial verification, the agent will distribute secret key pairs for the infrastructure.
\begin{equation}
\label{secretkey}
SecretKey_{Inf_{i}}=\left \{ PK,SK \right \}.
\end{equation}

In equation \ref{secretkey}, $PK$ represents the public key of infrastructure $Inf_{i}$, which is equivalent to a public identity. All users can know the infrastructure through the public key, while $SK$ represents the secret key, which is only known by infrastructure $Inf_{i}$ itself and will not be disclosed to the public, so it is private. At the same time, the infrastructure will receive its own exclusive certificate and wallet address. The certificate is for the only identification device, and the wallet is for storing rewards.

\subsection{Message transmission}
\begin{figure}[htbp] 
\centering
\includegraphics[width=1.0\columnwidth]{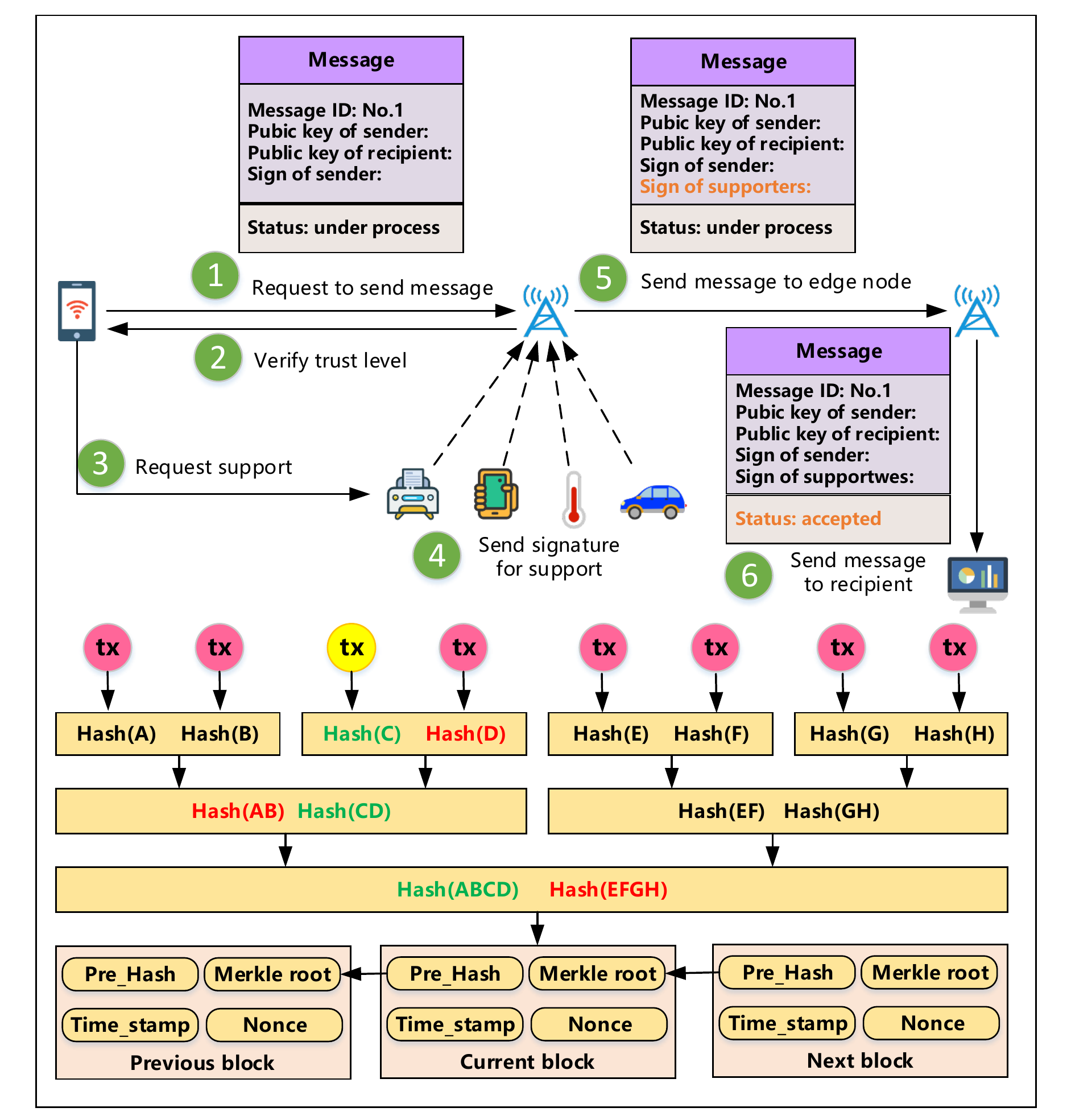}
\caption{The process of message propagation.}
\label{4}
\end{figure}

The process of message propagation in the framework is shown in Fig. \ref{4}. In the delivery process, we divide the infrastructure into three categories: message sender $Sender$, message receiver $Recerver$, and supporter $Supporter$. The infrastructure can change roles according to its own needs. When the infrastructure wants to send messages, its role is the sender. The infrastructure receiving messages becomes the receiver. The infrastructure proving the credibility of the sender in the process of sending is the supporter. In Fig. \ref{4}, for example, three roles and edge nodes are ready. First of all, the sender prepares the message he needs to send and uploads it to the edge node, equation \ref{message} represents the message structure.
\begin{equation}
\begin{split}
\label{message}
Message=\left \{ content|| PK_{sender}|| PK_{receiver}||\right.\\
\left.Signature_{sender}\right \},
\end{split}
\end{equation}

\begin{equation}
\label{status}
stutus=\left \{ -1, 0 ,1 \right \}.
\end{equation}

At first, the message includes the message content, the public key of the sender, the public key of the receiver, the signature of the sender. Each message has three statuses: being processed, received and failed to accept, as shown in equation \ref{status}, -1 represents failed to accept, 0 represents under processing, and 1 represents acceptance success. After receiving the sender's request, the edge node first authenticates the credit of sender and receiver. In this paper, the trust degree of almost every facility can be calculated by the following formula.
\begin{equation}
\label{trust}
T\left ( Inf_{i} \right )=\frac{\alpha l\left ( Inf_{i} \right )-\beta h\left ( Inf_{i} \right )}{\sum num\left ( message \right )}.
\end{equation}

This formula represents the trust degree of each infrastructure, and its result depends on the legal messages sent by each infrastructure and the number of paid messages. In equation \ref{trust}, $l\left ( Inf_{i} \right )$ represents the number of legal messages sent by each infrastructure, $h\left ( Inf_{i} \right )$ represents the number of illegal messages sent by each infrastructure, $\alpha$ and $\beta$ are the standard parameters, and $\sum num\left ( message \right )$ represents almost all messages sent by the infrastructure.
The credit of the sender and receiver can be represented by equation \ref{cre}.
\begin{equation}
\label{cre}
Cre(Inf_{i})=\gamma \cdot T(Inf_{i})+\delta \cdot q(Inf_{i}).
\end{equation}

In this formula, $\gamma$ and $\delta$ are harmonic coefficients, $T(Inf_{i})$ represents the trust degree of infrastructure, and $q(Inf_{i})$ represents the task quality assessment of edge nodes to infrastructure.
If the sender's credit rating meets the requirements, the edge node will collect the signature of the message's certifier. At this time, the sender needs to send a request to the surrounding users with higher credit rating, and reward them with a certain reward. After the supporter receives the task, it will send the signature to the edge node.
\begin{equation}
\begin{split}
\label{message2}
Message=\left \{ content|| PK_{sender}|| PK_{receiver}||\right.\\
\left.Signature_{sender}||Signature_{supporter}\right \}.
\end{split}
\end{equation}

The final qualified message is shown in equation \ref{message2}, which adds the signature of the supporters compared with the original message. When the number of supporters meets the requirements, the edge node will find the community of the receiver, and then send the information to the edge node of the community, and then the edge node will send the information to the recipient.

\subsection{Elect bookkeeper}
The edge nodes in the architecture can communicate and compete with each other. When there are various kinds of messages in the system, each edge node wants to compete for the position of bookkeeper, because it can get higher reward. This part introduces how to elect bookkeeper through consensus mechanism.

In bitcoin transaction, the problem is usually solved by pow. In this paper, we propose a POR consensus mechanism. Firstly, we define the problem solving formula, as shown in equation \ref{por}.
\begin{equation}
\label{por}
Hash(nonce|Hash(blockhead))\leq D\left ( Q \right )\cdot target.
\end{equation}

The process of generating a block is to find a nonce, so that the hash value is less than or equal to a specified target pre value, even if equation \ref{por} holds. There are many fields in the block header, one of which is the random number nonce that we can set, and the other is the hash value of the original block header. The process of solving is to try the random number continuously until we find the appropriate nonce value. Where $D\left ( Q \right )$ is the difficulty coefficient of the solution. The difficulty of solution is a very key parameter. When the difficulty of solution is too small, the block out time will be too short, which is easy to cause block chain bifurcation and threat to the system. When the difficulty of solution is too large, it will reduce the enthusiasm of the solver and cause a vicious circle. Therefore, the difficulty of solution should be adjusted according to the actual situation. The adjustment formula is as follows:
\begin{equation}
\label{diff}
D(Q)=\frac{{target_{D(Q)=1}}}{target},
\end{equation}
where, the denominator represents the current target, and the numerator represents the target value when the difficulty is 1 (when the difficulty is 1, it means the lowest difficulty).
Among them, target adjustment also needs to follow certain rules, and its adjustment formula is as follows:
\begin{equation}
\label{target}
target=target\times \frac{actual\; time}{expected\; time}.
\end{equation}

In equation \ref{target}, actual time refers to the actual time needed to generate a block, expected time refers to the expected time to generate a block, and the value of target changes continuously with the formula.

Bookkeeper selection is based on the trust and calculation ability of each bookkeeper. In the voting stage, each edge node with voting right can vote for the edge node it trusts. The weight of voting depends on the trust degree of the edge node.
The credit of edge nodes can be obtained by equation \ref{cre2}.
\begin{equation}
\label{cre2}
Cre(Inf_{j})=\frac{1}{1+e^{-T(Inf_{j})}}.
\end{equation}

The higher the credit, the higher the proportion of voting. At the end of voting, the edge node with the highest number of votes becomes the whole node, and the other edge nodes are the light nodes. All the nodes will run for bookkeeper through workload proof, which is to solve a nonce value together, so that the value of hash is equal to the set target value. The difficulty of the problem is adjusted with the solution speed of the node. The first node to solve the problem will get the bookkeeping right. The speed of solving the problem generally depends on the computing power of the node. The higher the computing power, the faster the solution speed and the greater the probability of success. Therefore, the node with higher computing power is more likely to become bookkeeper. The bookkeeper's responsibility is to collect all behavior data in the IoT, arrange them in chronological order, generate Merkle trees for all things according to hash values, store them in the block body of the block, and connect the block to the longest legal chain. The process is shown in Fig. \ref{4}. Therefore, the block contains the list of things, and the block header contains the hash value of the previous block and other information.

\subsection{Verification and reward}
In many scenarios, the existence and accuracy of messages are challenged. Therefore, we propose a method to verify the existence of messages: backtracking. The success of this method depends on the application of Merkle tree. Merkle tree can provide Merkle proof, that is, path from root back to edge. Taking Fig. \ref{4} as an example, suppose a light node wants to know whether the yellow transaction in the figure is included in the Merkle tree. The light node does not contain the transaction list, does not have the specific content of this Merkle tree, and only has a root hash value. At this time, the light node sends a request to a whole node to prove that the yellow transaction is included in the Merkle proof in this Merkle tree. After the whole node receives the request, it only needs to send the three hash values marked in red to the light node. With these hash values, the light node can locally calculate the three green hash values marked in the graph. First, calculate the hash value of the yellow transaction, that is, the green hash value directly above it, and then connect it with the red hash value next to it to calculate the green hash value of the upper node. Then, we can splice, calculate the upper green hash value, and then splice, we can calculate the root hash value of the whole tree. The light node compares the root hash value with the root hash value in the block header. If it is the same, it proves that the yellow transaction is in the Merkle tree.

\begin{figure}[H] 
\includegraphics[width=1.0\columnwidth]{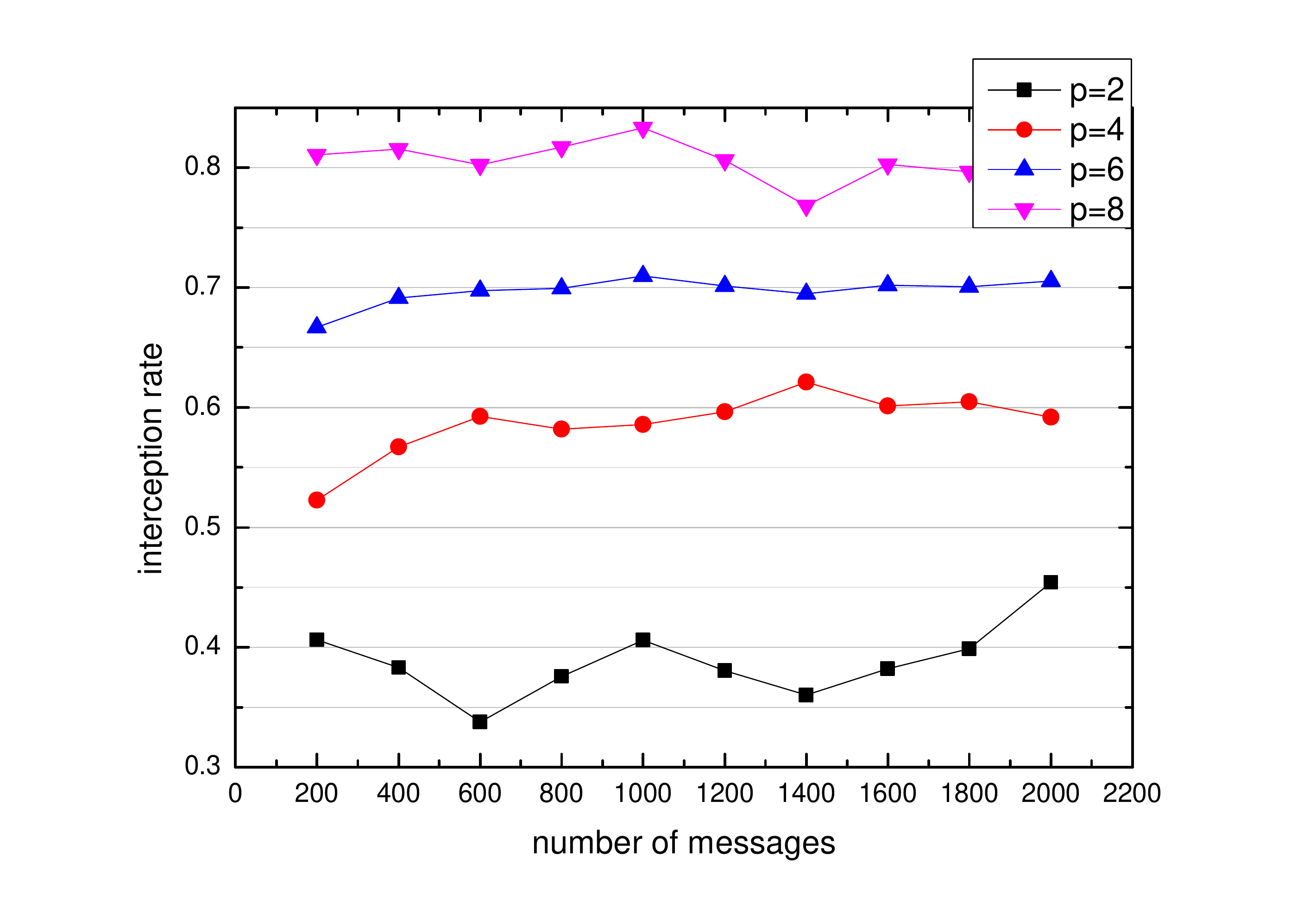}
\includegraphics[width=1.0\columnwidth]{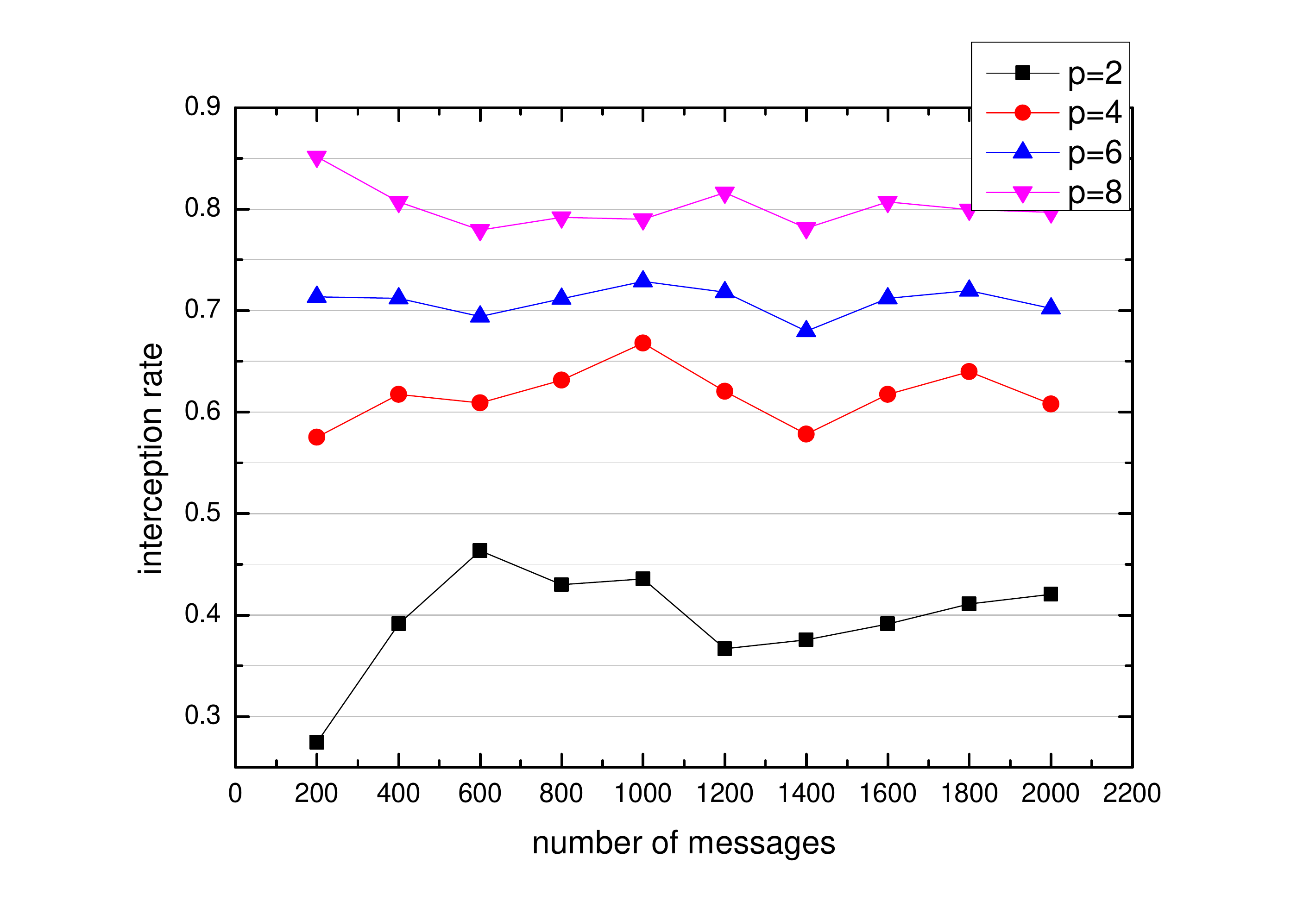}
\includegraphics[width=1.0\columnwidth]{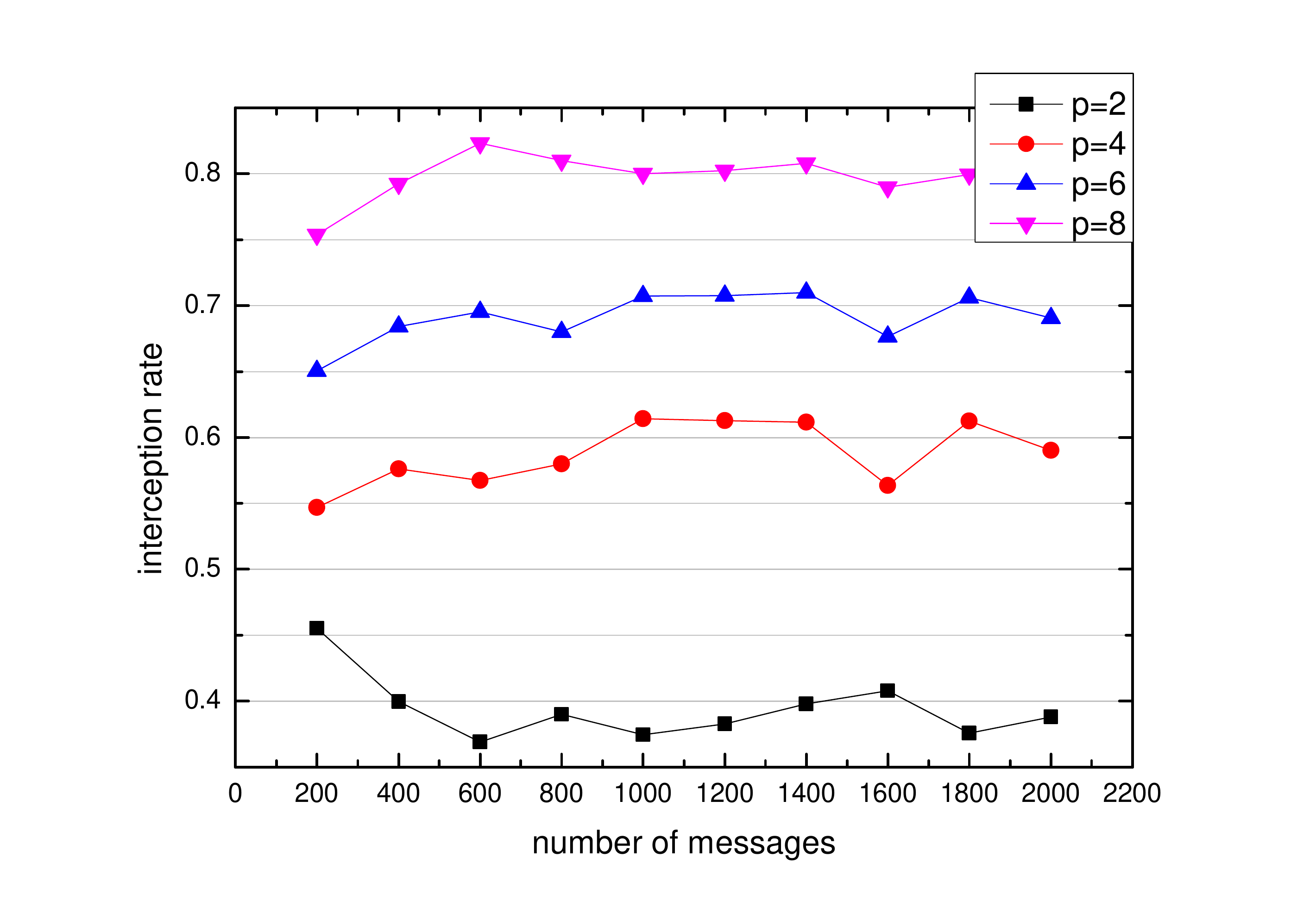}
\caption{Message interception rate under different attack rates.}
\label{5}
\end{figure}

When the message transmission is successful and has been connected to the blockchain, the bookkeeper will receive a reward and update its trust value. We introduce the mining pool model in the blockchain into reward distribution. In this model, the edge node can be regarded as the miner, and the infrastructure managed by him can be regarded as the miner. When the miner successfully generates the block and connects to the blockchain, the reward obtained will be regarded as the income of the whole mining pool. Miners will receive 50\% of their income, and the remaining income will be distributed according to the principle of "more work, more pay" for each infrastructure. The contribution of infrastructure is determined by the number of messages it sends. The more reliable messages it sends, the higher the profit it gets from dividends. The infrastructure can also gain revenue by testifying to the sender, and finally update the trust of the infrastructure.

\section{Simulation Experiment and Analysis}
In this chapter, we prove the performance of the proposed algorithm through simulation experiments. The simulation experiment is carried out on the operating system of windows 10, the code runs on eclipse, the program is written in Java language, and the final experimental results are displayed by the origin drawing tool. TABLE \ref{tbl} shows the range of parameters in the experiment.

\begin{table}[htpb]
\caption{The Settings of Parameters}
\label{tbl}
\begin{tabular}{lm{3cm}cm{2cm}}
\hline
Parameter Items                      & The Range    \\ \hline
the amount of infrastructures      & 100            \\
the amount of edge nodes           & 20         \\
the amount of communities              & 20   \\
the connection rate of infrastructures    & 0.6      \\
the trust level of infrastructures        & U[1,10] \\
the trust level of edge nodes           & U[1,10]\\
parameter $\alpha$               & 0.5          \\
parameter $\beta$                & 0.5          \\ \hline
\end{tabular}
\end{table}

\begin{figure}[htbp] 
\includegraphics[width=1.0\columnwidth]{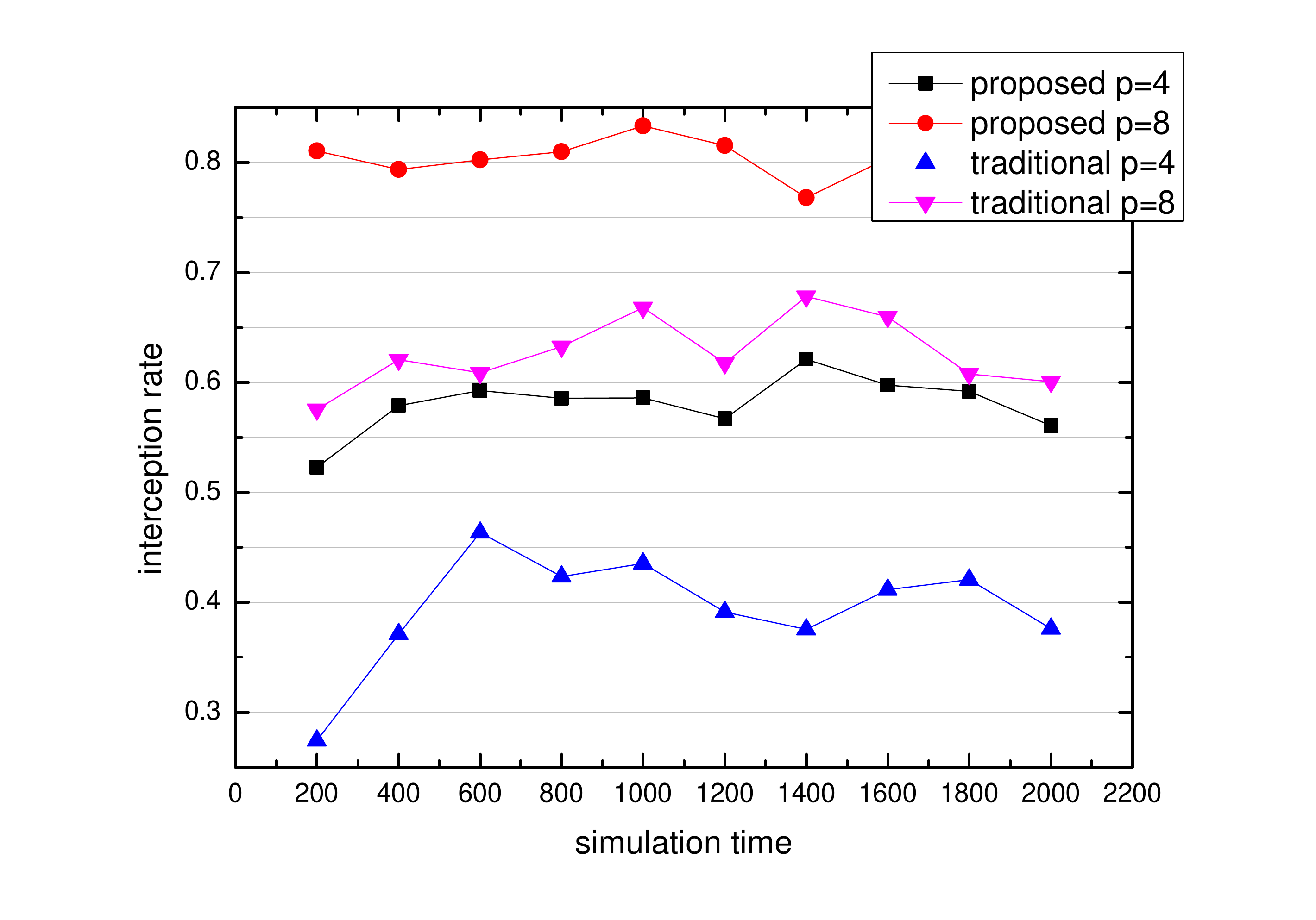}
\caption{The comparison of message interception rate between proposed and traditional scheme.}
\label{6}
\end{figure}

Fig. \ref{5} shows the interception rate of our scheme under different attack rates. Message interception rate refers to the ratio of the number of messages successfully intercepted to all illegal messages. The first figure in Fig. \ref{5} shows the interception rate of the method under different witnesses when the illegal attack rate is 0.2. The second figure and the third figure show the situation when the illegal attack rate is 0.4 and 0.6 respectively. From the figure, we can see that when p represents the number of supporters, when the number of supporters is 2, the message interception rate of the architecture is about 0.4, when the number of supporters is 4, the message interception rate is about 0.5 to 0.6, when the number of supporters is 8, the message interception rate is about 0.8. Experiments show that the interception rate of our proposed architecture increases with the increase of the number of supporters. Therefore, when the demand of supporters is large, the proof of message reliability will be more accurate, which can intercept illegal messages and prove legitimate messages. Therefore, our proposed mechanism can improve the reliability of message transmission.

Fig. \ref{6} compares the proposed scheme with the traditional centralized scheme. As can be seen from the figure, the scheme proposed in this paper has a higher interception rate than the traditional scheme when the supporter is 4, and a higher interception rate than the traditional scheme when the supporter is 8. But when the supporter is 4, the interception rate of the proposed scheme is lower than that of the traditional scheme when the supporter is 8, which shows that the number of supporters has a strong decisive role. At the same time, it can be seen that the scheme proposed in this paper has more obvious interception effect than the traditional centralized method in the same situation. In conclusion, the blockchain based scheme proposed in this paper has obvious advantages in ensuring transmission reliability.

\section{Conclusion}
The security of data transmission in the Internet of things has been widely concerned. With the rapid growth of Internet of things devices and data, the traditional cloud centric centralized processing mode is facing overload. The emergence of edge computing can alleviate the trouble of data centralized processing to a certain extent, but it still can not fundamentally solve the drawbacks of cloud object architecture. As a new technology, blockchain plays an important role in many fields. This paper combines blockchain with Internet of things and edge computing, and proposes a message transmission architecture based on blockchain. This architecture abandons the traditional cloud server, adopts the distributed architecture, and allocates the cloud processing to each participant. The solution of blockchain can make the data have the advantages of non-tamper ability and traceability. Through simulation experiments, it can be proved that the architecture proposed in this paper can ensure the transmission of messages in the Internet of things, and improve the reliability of transmission.

In the future work, we will always pay attention to the latest research progress, and intend to combine edge computing with network virtualization to conduct more in-depth research.
\bibliography{references}
\end{document}